\documentclass[twocolumn]{aastex63}
\usepackage{enumitem}

\shorttitle{Be star discs}
\shortauthors{Nixon \& Pringle}
\begin{document}
\title{Be star discs: powered by a non-zero central torque}
\author[0000-0002-2137-4146]{C.~J.~Nixon}
\affiliation{School of Physics and Astronomy, University of Leicester, Leicester, LE1 7RH, UK}
\author{J.~E.~Pringle}
\affiliation{School of Physics and Astronomy, University of Leicester, Leicester, LE1 7RH, UK}
\affiliation{Institute of Astronomy, Madingley Road, Cambridge, CB3 0HA, UK}

\email{cjn@leicester.ac.uk}

\begin{abstract}
Be stars are rapidly rotating B stars with Balmer emission lines that indicate the presence of a Keplerian, rotationally supported, circumstellar gas disc. Current disc models, referred to as ``decretion discs'', make use of the zero torque inner boundary condition typically applied to accretion discs, with the `decretion' modelled by adding mass to the disc at a radius of about two per cent larger than the inner disc boundary. We point out that, in this model, the rates at which mass and energy need to be added to the disc are implausibly large. What is required is that the disc has not only a source of mass but also a continuing source of angular momentum. We argue that the disc evolution may be more physically modelled by application of the non-zero torque inner boundary condition of \cite{Nixon:2020aa}, which determines the torque applied at the boundary as a fraction of the advected angular momentum flux there and approaches the accretion and decretion disc cases in the appropriate limits. We provide supporting arguments for the suggestion that the origin of the disc material is small-scale magnetic flaring events on the stellar surface, which, when combined with rapid rotation, can provide sufficient mass to form, and sufficient angular momentum to maintain, a Keplerian Be star disc. We discuss the origin of such small-scale magnetic fields in radiative stars with differential rotation. We conclude that small-scale magnetic fields on the stellar surface, may be able to provide the necessary mass flux and the necessary time-dependent torque on the disc inner regions to drive the observed disc evolution.
\end{abstract}
  

\section{Introduction}
Be stars are rapidly rotating, main sequence stars that somehow manage to form low density, equatorial, circumstellar discs in Keplerian rotation. These circumstellar discs of gas, that are the origin of the defining Balmer lines in Be stars, are generally seen to be variable, and sometimes to come and go \citep[see, for example, the reviews by][]{Porter:2003aa,Owocki:2006aa,Rivinius:2013aa,Okazaki:2016aa}. Not all rapidly rotating B stars display circumstellar discs; for example, the Bn stars are as numerous as the Be stars and show rapid rotation when seen equatorially through shallow and broad atmospheric lines, but no Balmer emission. We can define the break-up, or maximal, angular velocity of the star as $\Omega_{\rm K} = (GM_\star/R_\star^3)^{1/2}$, where $M_\star$ is the stellar mass and $R_\star$ the stellar radius.\footnote{Note that a rapidly rotating star is strongly distorted with the equatorial radius being larger than the polar one.  Thus the definition of $R_\star$ is somewhat uncertain. We gloss over this difficulty here, but refer the reader to \cite{Porter:2003aa} and \cite{Rivinius:2013aa} for a fuller discussion.}  Then a typical Be star has an angular velocity of $\left<\Omega_\star\right> \approx 0.8\,\Omega_{\rm K}$ \citep{Porter:2003aa,Rivinius:2013aa}. 

In this Letter, we concern ourselves with the formation, variability and disappearance of the circumstellar disc material. In Section~\ref{current},  we discuss current models for these variations. We show that  the assumptions made in these models are not physically reasonable, in terms of the magnitudes of the required fluxes of mass, energy and angular momentum. In Section~\ref{discformation}, we discuss current ideas for disc formation -- hydrodynamic and magnetic -- and argue strongly for the latter, in terms of variable, small-scale, equatorial magnetic fields. In Section~\ref{magfields} we discuss the origins of such fields and why they are likely only to be found in objects such as Be stars. We conclude in Section~\ref{conclude}.

\section{Current models of Be star disc variability}
\label{current}
Models for the time-variability of Be stars discs, and their use in modelling observations, especially of $\omega$ CMa, are provided by a number of authors \citep{Hanuschik:1993aa,Carciofi:2012aa,Haubois:2012aa,Ghoreyshi:2017aa,Rimulo:2018aa,Ghoreyshi:2018aa}. In these the modelling of the disc variations proceeds as follows:

\begin{enumerate}

\item The disc is assumed to have an inner boundary, $R_{\rm in}$, close to the stellar surface, $R_\star$, which has a zero-torque boundary condition ($f=0$ in the notation of \citealt{Nixon:2020aa} -- see Section~\ref{finite} below).~\footnote{This means that, contrary to the impression given in these papers, these discs are not {\em bona fide} decretion discs \citep{Pringle:1991aa,Nixon:2020aa}.}

\item The disc is assumed to have an outer boundary at which the disc sound speed is approximately equal to the escape velocity, typically at around $R_{\rm out} \approx 400 - 1,000 \, R_\star$. There the boundary condition is also zero-torque, so that the disc loses mass freely at that radius.

\item At the start of an outburst event, mass is added to the disc at some rate $\dot M$, at a radius $R_{\rm add} =  R_{\rm in} (1+ \epsilon)$ which is very close to the inner radius with typically $\epsilon \approx 0.02$. The matter is added to the disc with the Keplerian velocity for that radius. For the decline phase of an event, $\dot M$ is set to zero, and (almost) all of the disc drains through the inner boundary back onto the star.

\end{enumerate}

These models imply that the rate at which mass is added to the disc greatly exceeds the rate at which matter finds its way into the observable circumstellar disc. For example, in a steady disc with a zero torque inner boundary, the fraction of added mass that finds its way into the circumstellar disc is $\approx 0.5 \epsilon (R_{\rm in}/R_{\rm out})^{1/2} \approx 3 \times 10^{-4}$ \citep{Nixon:2020aa}. This agrees with the estimate given by \citet[][see also \citealt{Ghoreyshi:2018aa}]{Rimulo:2018aa}. During the growth phase, \cite{Rimulo:2018aa} find that the disc is required to grow at a rate ${\dot M}_{\rm disc} = 10^{-9}\,M_\odot$/yr, which can be comparable to a typical mass loss rate in the stellar wind \citep{Krticka:2014aa}\footnote{Here we use the stellar parameters given by \cite{Okazaki:2016aa} for a typical B0 main sequence star.}. This implies that typically mass is added to the disc at a rate ${\dot M} \sim 3\times 10^{-6}\,M_\odot$/yr,  which is around $3\times 10^3$ times the stellar wind mass loss rate. For a typical Keplerian velocity of $\approx 700$\,km/s this requires an energy input of several per cent of the stellar luminosity. Moreover, because the addition of mass implies the addition of angular momentum, it is apparent that adding mass, and therefore angular momentum, at these rates would imply an unrealistically large flux of angular momentum (i.e.\@ torque) at the stellar surface.

We conclude that while the current models provide a simple means of modelling the Be star discs at radii away from the stellar surface, as they stand, they also contain hidden assumptions which are not physically plausible. 

\section{Models for disc formation}
\label{discformation}

Rivinius and colleagues \citep{Rivinius:2013aa}, in an extensive and authoritative review article, provide a discussion and historical overview of the many mechanisms that have been proposed for the formation and variability of discs in Be stars. \cite{Rivinius:2013aa} conclude that the current best options for providing sources of mass and angular momentum to the circumstellar Be star discs are {\em either} (i) material launched hydrodynamically from the stellar surface by non-radial pulsations, {\em or} (ii) material launched by {\em small-scale} magnetic fields.~\footnote{There is a long history of discussion of disc formation by large-scale (quasi-dipole) fields, but \cite{Rivinius:2013aa} rule these out on both theoretical and observational grounds (see also Section 4).}

\subsection{Hydrodynamic disc launching}
\label{hydro}

The attraction of tying the launching of Be star discs to the observed non-radial pulsations is that there seems to be a strong correlation between the disc launching and variability and the presence and detailed behaviour of such pulsations \citep[for example][]{Baade:2016aa,Baade:2018aa,Semaan:2018aa,Neiner:2020aa}. In Section~\ref{current} we have detailed the physical requirements for the launching and maintenance of a Be star disc in terms of the current models. \cite{Owocki:2006aa} explains why such requirements cannot be met by a purely hydrodynamic mechanism (such as disc launching by non-radial pulsations). The fundamental, and insuperable, problem is that the velocity of the launched material, relative to the stellar surface, typically exceed the sound speed at the stellar surface by factors of $\sim 5 - 30$. Furthermore, as we have seen, unless the launching mechanism can also provide a continuous supply of angular momentum the mass and energy fluxes within the star/disc interface are implausibly large. Providing such a supply of angular momentum by purely hydrodynamic means would involve hypersonic, severely dissipative fluid motions. Thus, a more plausible picture for the relationship between disc launching and maintenance and non-radial pulsations might be one in which the launching mechanism and/or presence of a disc causes the non-radial pulsations, rather than the other way round.\footnote{For example, if, as we argue below, disc launching comes about as a result of a peak in dynamo action, and if such a peak is able to lead to the excitation of non-radial pulsations.}

\subsection{Non-zero central torque accretion discs}
\label{finite}
Recently, \cite{Nixon:2020aa} have noted that a more appropriate inner boundary condition might be one in which the torque at the inner boundary is finite, but {\em non-zero}. They show the properties of both steady and time-dependent discs for which there is an angular momentum source (torque) at the inner edge which is a factor $f$ times the advective angular momentum sink there. In this notation, standard accretion discs \citep[e.g.][]{Pringle:1981aa} correspond to $f=0$, and standard decretion discs \citep[e.g.][]{Pringle:1991aa} correspond to the limit $f \rightarrow \infty$.  In the Be star case, a source of angular momentum ($f > 0$) can be achieved if the inner disc boundary is now at the Alfv\'en  radius, $R = R_M$ \citep{Pringle:1972aa},  which is determined by the radius at which the disc is being truncated by small-scale magnetic fields \citep[cf.][]{Rivinius:2013aa}. We then require that radius, $R_M$, to be greater than the co-rotation radius $R_\Omega$ at which the stellar angular velocity equals the disc Keplerian velocity. If we define the break-up angular velocity of the star as $\Omega_K = (GM_\star/R_\star^3)^{1/2}$, then for a typical rotation velocity of $\Omega_\star \approx 0.8 \Omega_K$ \citep{Porter:2003aa,Rivinius:2013aa} we find that $R_\Omega/R_\star \approx (\Omega_K/\Omega_\star)^{2/3} \approx 1.16$. For a typical inner disc density of $\rho_0 \sim 10^{-10} - 10^{-12}$ g/cm$^3$ \citep{Silaj:2010aa}, and temperature of $T_0 \sim 15,000 - 20,000$ K, the field strengths required to ensure that $R_M \ge R_\Omega$ would be around $B \sim 5 - 50$ G, on radial scales of around $\sim 0.2 R_\star$. Such fields are not currently observable \citep[for example][]{Wade:2016aa}.

If these conditions could be achieved, then more physically plausible, time-dependent disc models, essentially equivalent to those described in Section~\ref{current}, could operate as follows:

\begin{enumerate}
\item In the disc growth phase, matter (and angular momentum) is added to the disc at some radius $R_{\rm add} > R_\Omega$. The inner disc boundary is such that a continuing (magnetic) torque is provided to the disc material. In the parameterization of \cite{Nixon:2020aa} this implies $f \gg 1$~\footnote{Note that simply increasing the value of $f$ from $f=0$ (standard accretion) to $f \gtrsim 1$ at the inner boundary is likely not sufficient as significant accretion can then still continue, despite the net flow of angular momentum being reversed \citep[cf.][]{Popham:1991aa}. Note too that a true decretion disc occurs only in the limit $f \rightarrow \infty$ \citep{Nixon:2020aa}.}. It would be plausible for the matter to be provided to the disc by the same magnetic processes that provide the inner torque (like solar coronal mass ejections, but on a much reduced scale, or a small-scale multipole version of the “slingshot prominences” discussed by \citealt{VillarrealDAngelo:2018aa,VillarrealDAngelo:2019aa} and \citealt{Jardine:2019aa}).
\item In the decline phase, magnetic activity declines until $R_M < R_\Omega$, decreasing the rate at which mass is added to the disc, and removing the inner magnetic torque (so that now $f \ll 1$).
\end{enumerate}

 We have argued above that in order for most of the material launched into the disc to remain there (at least initially) it is necessary for the launching mechanism to be able to provide a continuing source of angular momentum. Thus, we conclude that the existence and the evolution of the disc is most plausibly provided by the time-dependent behaviour and activity of small-scale magnetic fields close to the equatorial stellar surface.

\section{Magnetic fields}
\label{magfields}
The idea that the inner parts of Be star accretion discs are controlled by small-scale magnetic fields is not new. \cite{Smith:1989aa} interpreted the transients and ejections in $\lambda$ Eri in terms of small-scale magnetic flaring processes similar to those on the Sun, T Tauri stars, and several other types of magnetically active cool stars. He suggests that ``quasi-cycles'' in at least some classical Be stars find their origin in chaotic knots of magnetic field that periodically drift to the surface, as on the Sun, and dissipate through flaring triggered by sudden changes in field topology, perhaps jostled by surface velocity fields provided by, for example, their nonradial pulsations. Further, \cite{Smith:1994aa,Smith:1997aa} suggest that their observations of variability in $\lambda$ Eri can be interpreted as being due to magnetic loops and magnetically induced flares. More recently \citep{Smith:2016aa} he has applied such ideas to the X-ray production in $\gamma$ Cas. \citet[][see also \citealt{ud-Doula:2002aa}]{Owocki:2006aa} consider large-scale dipole magnetic fields as a means of providing torques on an outflowing channeled wind and conclude that this mechanism is not viable. \cite{Rivinius:2013aa} note that no magnetic fields have yet been detected on any Be star, and suggest an upper limit to a ordered, net, line-of-sight field component of around 100\,G. They also discuss the evidence for small-scale magnetic activity as a explanation for rapid line variability in the optical and the UV. 

The question then is: what is the source of such small-scale magnetic fields? And why do they only appear in Be stars? The presence of large-scale magnetic fields in a subset of early-type stars (for example the Ap/Bp stars) is usually ascribed to processes occurring at late stages of the formation process, for example a late merger which induces strong differential rotation \citep{Ferrario:2009aa,Jermyn:2020aa}. But ``conventional wisdom'' suggests that early-type stars, lacking outer convective zones, cannot produce active magnetic regions \citep[see, for example,][]{Zinnecker:1994aa,Zinnecker:1994ab}. This conclusion was questioned by \cite{Tout:1995aa} who argued that, even in a purely radiative star, dynamo activity could be driven by differential rotation combined with the effects of buoyancy (Parker instability) on the sheared seed fields. This idea was followed up in more detail by \cite{Spruit:2002aa} and by \cite{Braithwaite:2004aa}. Recent MHD simulations of the interaction between differential rotation and magnetic fields in radiatively stable fluids are presented by \cite{Simitev:2017aa}, and by \cite{Jouve:2020aa}. In an investigation of spots on radiative A- and B-stars, \cite{Balona:2019aa} concludes that in such stars ``differential rotation may be sufficient to create a local magnetic field via dynamo action''. 

Thus the next question is: why should these stars possess sufficient {\em differential} rotation to power dynamo action? Both Be stars and Bn stars are known to rotate rapidly, although there is a value of the rotation rate above which all the rapid rotators are Be stars \citep{Rivinius:2013aa}.  Most modelling of the shape and structure of these rapidly rotating stars is usually made using the assumption of uniform rotation \citep[for example,][]{Rivinius:2013aa}. However, it is well established that stellar rotation, together with the requirements of local hydrodynamic and thermal equilibrium leads to secondary flows within the star, which in turn lead to redistribution of the stellar angular momentum distribution. In radiative stars, such flows tend to lead to differential rotation, with the rotation velocities peaking at the equator \citep{Tassoul:1982aa,Garaud:2002aa}. Differential rotation would also be the norm if, as suggested by \cite{Bodensteiner:2020aa} and \cite{El-Badry:2020aa}, Be stars are binary interaction products, spun-up by mass transfer. In the presence of differential rotation, all that is then required is an initial source of magnetic flux. Seed fields for a dynamo could be provided originally during the formation process, or latterly by subsurface convection zones \citep{Charbonneau:2001aa,Jermyn:2020aa}, and by MRI-driven dynamo action in the disc itself \citep[e.g.][]{Martin:2019aa}.  Thus we propose that the small-scale magnetic fields needed to power Be discs (both in terms of mass and in terms of angular momentum) are produced by small-scale dynamo action in a differentially rotating zone close to the stellar equator. On this hypothesis, it might also be these motions that give rise to the behaviour of the non-radial pulsation modes.~\footnote{It is worth noting here that material being re-accreted from the disc can act as a source not only of magnetic flux, but also as a source of shear that can act as a driving mechanism for non-radial stellar pulsations \citep{Papaloizou:1978aa,Papaloizou:1980aa}.}

In this picture the Bn stars would be those rapid rotators which {\em currently} have fields that are too weak to ensure $R_{\rm M} > R_\Omega$, so that disc formation can take place. We suggest that this might come about for a combination of reasons: (i) since Bn stars appear on average to rotate more slowly than Be stars it follows that Bn stars would need on average to have higher surface dynamo fields in order to launch a circumstellar disc, and (ii) stars which rotate more slowly would tend on average to have less efficient dynamos, thus to have dynamo cycles that spend more of the time in a low state, and so to spend more of the time unable to launch a disc.

\section{Conclusions}
\label{conclude}
We suggest that the variability of the circumstellar discs around Be stars should be modelled in terms of accretion discs with variable, but finite, central torque \citep{Nixon:2020aa}. We have argued that both the disc mass and the central torque are most likely provided by small-scale magnetic activity occurring close to the equator of the rapidly spinning Be star. We have put forward ideas as to how  magnetic activity can be initiated and maintained in rapidly rotating stars with radiative envelopes, and have suggested that in this picture it is the nature of the dynamo that gives rise to the distinction between Be stars and Bn stars.

\acknowledgments
We thank Rebecca Martin for useful correspondence. We thank the referee for a useful report. CJN is supported by the Science and Technology Facilities Council (grant number ST/M005917/1). This project has received funding from the European Union’s Horizon 2020 research and innovation program under the Marie Sk\l{}odowska-Curie grant agreement No 823823 (Dustbusters RISE project).

\bibliographystyle{aasjournal}
\bibliography{nixon}
\end{document}